%% file: ase2026_camera.tex
\documentclass[sigconf,screen]{acmart}

\AtBeginDocument{%
  \providecommand\BibTeX{{%
    Bib\TeX}}}

\usepackage{microtype}  
\usepackage{amsmath,amssymb}
\usepackage{algorithmic}
\usepackage{graphicx}
\def\BibTeX{{\rm B\kern-.05em{\sc i\kern-.025em b}\kern-.08em
    T\kern-.1667em\lower.7ex\hbox{E}\kern-.125emX}}
\usepackage{multirow}
\usepackage[table,xcdraw]{xcolor}
\usepackage{url}
\usepackage{booktabs}
\usepackage{subcaption}
\usepackage{xspace}
\usepackage{cleveref}
\usepackage{threeparttable}
\usepackage[many]{tcolorbox}
\usepackage{enumitem}

\usepackage{tcolorbox}
\usepackage{listings}
\usepackage{pifont}  

\DeclareRobustCommand{\mybox}[2][gray!20]{%
\begin{tcolorbox}[
        breakable,
        left=0pt,
        right=0pt,
        top=0pt,
        bottom=0pt,
        colback=#1,
        colframe=#1,
        width=\linewidth, 
        enlarge left by=0mm,
        boxsep=5pt,
        arc=0pt,outer arc=0pt,
        ]
        #2
\end{tcolorbox}
}

\newcommand{\lstbg}[3][0pt]{{\fboxsep#1\colorbox{#2}{\strut #3}}}
\definecolor{codegreen}{rgb}{0,0.6,0}
\lstdefinelanguage{diff}{
	frame=single,
	basicstyle=\ttfamily\scriptsize\bfseries,
	morecomment=[f][\color{red}]{---}, 
	morecomment=[f][\color{codegreen}]{+++},
	morecomment=[f][\lstbg{red!20}]{-},
	morecomment=[f][\lstbg{green!20}]{+},
	morecomment=[f][\color{blue}]{@@},
}

\usepackage{tikz}

\definecolor{OliveGreen}{rgb}{0,0.6,0}

\newcommand{\tool}{\textit{SPectre}\xspace}

\newcommand{\spdb}{$SP_{DB}$\xspace}
\newcommand{\sptax}{$SP_{Tax}$\xspace}
\newcommand{\speval}{$SP_{Eval}$\xspace}

\usepackage[font=small,skip=2pt]{caption}
\newcommand{\distance}{10pt}
\setcopyright{cc}
\setcctype{by}
\acmDOI{10.1145/3832783.3837455}
\acmYear{2026}
\copyrightyear{2026}
\acmISBN{979-8-4007-2882-2/2026/10}
\acmConference[ASE '26]{Proceedings of the 41st IEEE/ACM International Conference on Automated Software Engineering}{October 12--16, 2026}{Munich, Germany}
\acmBooktitle{Proceedings of the 41st IEEE/ACM International Conference on Automated Software Engineering (ASE '26), October 12--16, 2026, Munich, Germany}
\acmSubmissionID{ase26main-p694-p}
\received{2026-03-26}
\received[accepted]{2026-06-18}

\begin{document}
\abovedisplayskip=0pt
\abovedisplayshortskip=0pt
\belowdisplayskip=0pt
\belowdisplayshortskip=0pt
\title{One Is Not Enough: The Untold Story of Multiple Security Patches for One Vulnerability}


\author{Fangyuan Zhang}
\orcid{0009-0000-9599-1369}
\affiliation{%
  \department{College of Computer Science}
  \institution{Nankai University}
  \city{Tianjin}
  \country{China}
}
\email{fangyuanzhang@mail.nankai.edu.cn}

\author{Lyuye Zhang}
\correspondingauthor
\authornote{Lyuye Zhang and Lingling Fan are the corresponding authors.}
\orcid{0000-0003-3087-9645}
\affiliation{%
  \department{College of Cryptology and Cyber Science}
  \institution{Nankai University}
  \city{Tianjin}
  \country{China}
}
\affiliation{%
  \institution{Nanyang Technological University}
  \country{Singapore}
}
\email{zh0004ye@e.ntu.edu.sg}

\author{Lingling Fan}
\correspondingauthor
\authornotemark[1]  
\orcid{0000-0002-2428-9297}
\affiliation{%
  \department{DISSec, NDST, College of Cryptology and Cyber Science}
  \institution{Nankai University}
  \city{Tianjin}
  \country{China}
}
\email{linglingfan@nankai.edu.cn}

\author{Chengwei Liu}
\orcid{0000-0003-1175-2753}
\affiliation{%
  \department{College of Cryptology and Cyber Science}
  \institution{Nankai University}
  \city{Tianjin}
  \country{China}
}
\email{chengwei.liu@nankai.edu.cn}

\author{Yinan Li}
\orcid{0009-0004-5629-1377}
\affiliation{%
  \department{College of Cryptology and Cyber Science} 
  \institution{Nankai University}
  \city{Tianjin}
  \country{China}
}
\email{2120250736@mail.nankai.edu.cn}

\author{Liang Huang}
\orcid{0009-0008-8515-9673}
\affiliation{%
  \institution{Qi An Xin Technology Group}
  \city{Beijing}
  \country{China}
}
\email{huangliang@qianxin.com}

\author{Yang Liu}
\orcid{0000-0001-7300-9215}
\affiliation{%
  \institution{Nanyang Technological University}
  \country{Singapore}
}
\email{yangliu@ntu.edu.sg}

\author{Zheli Liu}
\orcid{0000-0002-2984-2661}
\affiliation{%
  \department{DISSec, NDST, College of Cryptology and Cyber Science}
  \institution{Nankai University}
  \city{Tianjin}
  \country{China}
}
\email{liuzheli@nankai.edu.cn}

\author{Sen Chen}
\orcid{0000-0001-9477-4100}
\affiliation{%
  \department{DISSec, NDST, College of Cryptology and Cyber Science}
  \institution{Nankai University}
  \city{Tianjin}
  \country{China}
}
\email{senchen@nankai.edu.cn}

\renewcommand{\shortauthors}{Zhang et al.}

\begin{abstract}
Security patches (SPs) are the main mechanism for fixing software vulnerabilities, yet a single vulnerability is not always resolved by a single patch: fixes may be completed incrementally, propagated across maintained branches, or replicated across related repositories. When patch records are incomplete, downstream users may observe only part of the required fix set and therefore apply only partial patching.
However, comprehensive patch discovery remains difficult because the prevalence and causes of the multi-SP phenomenon are still poorly understood. In this paper, we present the first large-scale empirical study of multi-SP vulnerabilities. By merging four major vulnerability databases, we construct a dataset of 6,053 multi-SP CVEs with 16,260 SPs, showing that 20.6\% of CVEs with patches involve multiple SPs and that merging databases increases recognized multi-SP CVE counts by 36-55\% over any single source. We further analyze why a vulnerability is associated with multiple SPs and derive a two-level taxonomy with 6 categories and 16 sub-categories. Based on these findings, we develop \tool, a taxonomy-driven prototype for comprehensive patch discovery.
On 300 multi-SP CVEs, after manually verifying ground-truth SPs,
\tool improves multi-SP patch coverage over representative patch-localization baselines, achieving 0.927 recall on same-repository cases and 0.873 recall on cross-repository cases after manual ground-truth verification.
On 100 recent CVEs recorded as single-patch by all public databases, \tool further discovers 28 previously unreported SPs across 20 CVEs. Our results show that multi-SP vulnerabilities are both prevalent and systematically underreported, motivating stronger patch-completeness awareness, improved vulnerability database curation, and relation-aware security tooling.
\end{abstract}

\ccsdesc[500]{Security and privacy~Software security engineering}
\ccsdesc[300]{Software and its engineering~Maintaining software}

\keywords{Software Security, Security Patch, Vulnerability Database}

\maketitle

\section{Introduction}

The prevalence of software vulnerabilities has grown exponentially with open-source ecosystems and dependency networks. 
Each newly disclosed vulnerability can propagate through thousands of downstream projects~\cite{liu2022demystifying}, amplifying its impact across the software ecosystem. 
Security patches (SPs), therefore, play a pivotal role in ecosystem-wide mitigation by embedding corrective logic that prevents exploitation and halts further vulnerability diffusion.

Fixing a vulnerability is rarely a one-time job. On the one hand, fixing can often be done incrementally: an initial patch may be incomplete or incorrect, and follow-up commits are needed to address overlooked cases, residual attack surfaces, or newly exposed exploitation vectors. For example, the Log4Shell vulnerability has four successive patches over 18 days before its exploitation vectors were fully addressed~\cite{cve202144228}. On the other hand, a vulnerability may affect multiple maintained versions or related repositories, and therefore require distinct patches in each of them. For instance, CVE-2022-0778~\cite{cve-2022-0778} was patched on several actively maintained OpenSSL branches, each with a different implementation context.
In both scenarios, the effective vulnerability fixing is reflected by multiple SPs rather than an isolated commit, also reported by prior studies~\cite{woo2025large}. When vulnerability databases record only part of that set, downstream users may obtain a partial view of the vulnerability fixing scope. Missing follow-up patches may lead to false belief that a vulnerability has been fully fixed when important attack surfaces remain exposed. Missing branch- or repository-specific patches can prevent maintainers from locating the applicable fix, delay patch deployment, or encourage incorrect patch porting across incompatible contexts. 
Therefore, improving the coverage of recorded SPs is essential for characterizing real-world vulnerability fixing.

Unfortunately, existing patch localization approaches do not effectively support comprehensive patch discovery for vulnerabilities. Prior studies largely formulate patch identification as commit-level relevance ranking~\cite{tan2021locating,wang2022vcmatch,dunlap2024vfcfinder,sabetta2024known,li2024patchfinder} or rely on link tracing across references and branches~\cite{xu2022tracking,sabetta2024known}. While these strategies may rank multiple SPs at high orders, no semantic relationships among SPs are modeled. 
SHIP~\cite{song2025not} moves beyond single-commit identification by predicting pairwise inter-commit relevance to cluster multiple SPs into a group. 
However, it models inter-commit relations with surface-level statistical signals, counts of shared code entities, cross-reference co-occurrence, and text similarity of commit messages, without reasoning about the deeper semantic dependencies.
Existing methods thus remain limited in discovering comprehensive patch sets for multi-SP vulnerabilities. A systematic understanding of why multi-SP vulnerabilities arise and how their SPs are related is therefore essential for improving patch coverage.

To address this gap, we conduct a large-scale empirical study of multi-SP vulnerabilities by merging four vulnerability databases, and develop a prototype tool, named \tool, to operationalize the resulting insights. Specifically, in \textbf{RQ1}, we quantify the prevalence of multi-SP vulnerabilities in existing vulnerability databases and measure how cross-source aggregation improves their coverage. In \textbf{RQ2}, we investigate why a single vulnerability is associated with multiple SPs by analyzing the semantic relationships among patches, and we construct a taxonomy of the reasons behind the multi-SP phenomenon, comprising 6 categories and 16 sub-categories. In \textbf{RQ3}, we translate these findings into \tool and evaluate whether the patterns and taxonomy findings can identify related SPs more effectively than existing patch localization baselines.
After manually verifying the ground-truth SPs,
\tool achieves 0.927 recall for the same-repository cases, and 0.873 recall in the cross-repository setting, recovering additional related SPs beyond representative ranking-based baselines at \textit{K}=50.
Finally, in \textbf{RQ4}, we assess whether \tool can discover previously unreported SPs for recently disclosed CVEs through manual verification and real-world validation, uncovering 28 unreported SPs, one of which has been officially acknowledged by GHSA~\cite{ghsa}.

In conclusion, we made the following contributions:
\begin{itemize}[leftmargin=20pt]
    \item We present the first large-scale empirical study of multi-SP vulnerabilities, integrating four major vulnerability databases, and quantify both their prevalence and the additional coverage obtained through cross-source aggregation.
    \item We provide a systematic semantic analysis of why a single vulnerability corresponds to multiple SPs , and derive a two-level taxonomy of the reasons behind the multi-SP phenomenon.
    \item We design \tool to operationalize these insights and demonstrate that it complements existing patch localization baselines by improving coverage of related SPs in multi-SP scenarios.
    \item We validate \tool on recently disclosed CVEs and uncover 28 previously unreported SPs after manual review, showing that databases systematically under-report multi-SP phenomenon.

\end{itemize}

\section{Background and Research Problem}

\textbf{Definitions.} For clarity, we introduce several key concepts used throughout this study:
\begin{itemize}[leftmargin=20pt]
  \item \textbf{Security Patch (SP):} A code commit that fixes the vulnerability, as often referenced in vulnerability databases~\cite{nvd,ghsa,xu2022tracking}.
  \item \textbf{Multi-SP Vulnerability:} A disclosed vulnerability explicitly associated with more than one SP in vulnerability databases.

\end{itemize}

\textbf{Research Problem.}
This study aims to systematically investigate the \textit{multi-SP phenomenon}, where multiple SPs correspond to a single vulnerability (CVE), and develop a prototype tool to discover the comprehensive SPs.
We seek to understand \textit{how prevalent} multi-SP vulnerabilities are in existing vulnerability databases, \textit{why} a single vulnerability requires multiple SPs, \textit{how} the empirical insights can support the identification of related SPs, and \textit{whether} unreported SPs for CVEs can be discovered in practice.

\section{Data Collection Methodology}

Although prior work~\cite{song2025not} collected a multi-SP dataset, it relies on two vulnerability databases, NVD and Snyk, and may therefore have limited coverage of SPs and multi-SP CVEs. To systematically study the multi-SP phenomenon at a larger scale, we construct \spdb by collecting SPs from four widely used and publicly accessible vulnerability databases. Since these databases share a common vulnerability identifier (CVE-ID), we merged the SP records associated with the same CVE across databases with de-duplication.

We leveraged four well-established vulnerability databases used in recent work~\cite{li2024patchfinder,xie2024unveiling,wu2024vision} to obtain SPs with CVEs: NVD~\cite{nvd}, Snyk Vulnerability Database~\cite{snykdb}, GHSA~\cite{ghsa}, and Google Open Source Vulnerabilities (OSV) database~\cite{osv}. 
We crawled CVE entries and associated references from 2005 to 2026, and applied regular expressions to extract links that both contain the keyword ``commit'' and a commit hash, following the previous works~\cite{tan2021locating,li2024patchfinder}.

The detailed data collection methodology is as follows:
\ding{172} We collected the CVEs with SPs from four major vulnerability databases.
\ding{173} We merged the extracted SPs from all four vulnerability databases on a per-CVE basis with de-duplication. In total, we collected 29,445 CVEs associated with 39,659 SPs.
\ding{174} To study the multi-SP vulnerabilities, we filtered out CVEs with one SP and retained only those linked to more than one distinct SP. This process yielded 6,053 CVEs (20.6\% of total collected CVEs), associated with 16,260 SPs, constituting a multi-SP dataset as \textbf{\spdb} from the merged SP set of four major vulnerability databases. \spdb serves as the foundation for the subsequent analysis in research questions.

\section{Empirical Study}
We organize our empirical study around the following RQs:

\begin{itemize}[leftmargin=20pt]
  \item \textit{RQ1: How prevalent are multi-SP vulnerabilities in existing vulnerability databases?} 
  \item \textit{RQ2: What are the reasons behind the multi-SP phenomenon for a single vulnerability?}
  \item \textit{RQ3: To what extent can our empirical findings support the effective identification of multiple SPs?}
  \item \textit{RQ4: Can \tool find unreported SPs for the recent CVEs?}
\end{itemize}

\begin{figure}[t]
  \centering
  \includegraphics[width=\columnwidth]{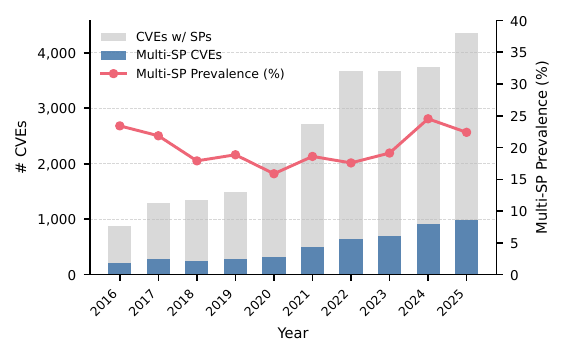}
  \caption{Annual prevalence of multi-SP phenomenon.}
  \label{fig:annual_prevalence}
\end{figure}

\subsection{RQ1: Prevalence of Multi-SP Vulnerabilities}
\label{sec:prevalence}

\subsubsection{Prevalence of Multi-SP Vulnerabilities in \spdb}
\label{sec:db_study}

\begin{figure}[t]
  \centering
  \includegraphics[width=\columnwidth]{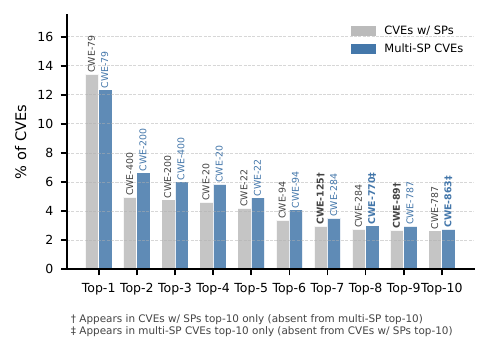}
  \caption{Top-10 CWE distribution in CVEs w/ SPs vs.\ multi-SP CVEs.}
  \label{fig:cwe_distribution}
\end{figure}

\begin{figure}[t]
  \centering
  \includegraphics[width=\columnwidth]{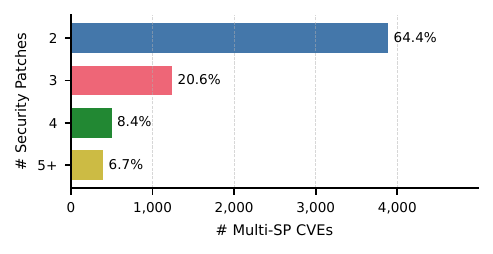}
  \caption{Distribution of SP counts per multi-SP CVE.}
  \label{fig:sp_distribution}
\end{figure}

\Cref{fig:annual_prevalence,fig:cwe_distribution,fig:sp_distribution} present three views of multi-SP prevalence in \spdb: the annual prevalence of multi-SP phenomenon, the differences in vulnerability types between all CVEs with SPs and multi-SP CVEs only, and the distribution of patch counts per multi-SP CVE.

\noindent\textbf{Annual trend.}
As shown in \Cref{fig:annual_prevalence}, multi-SP CVEs consistently account for 15-25\% of all CVEs with security patches throughout 2016-2025, demonstrating that multiple SPs per vulnerability is a systemic and persistent phenomenon. More notably, prevalence grows steadily from 15.9\% in 2020 to 24.5\% in 2024, suggesting that the multi-SP phenomenon has been intensifying in recent years.

\noindent\textbf{CWE distribution.}
To examine whether certain vulnerability types are more prone to the multi-SP phenomenon,
we compare the top-10 CWE distributions of all CVEs with SPs and of multi-SP CVEs (\Cref{fig:cwe_distribution}). The two lists share 8 of 10 CWEs. CWE-770 (Allocation of Resources Without Limits or Throttling) and CWE-863 (Incorrect Authorization) enter the multi-SP top-10 but are absent from the general top-10.
This may reflect their structural complexity:
resource throttling must be enforced at multiple architectural layers (e.g., API gateway, middleware, and application logic)~\cite{zargar2013survey}. Authorization decisions are distributed across modules and are highly context-dependent~\cite{son2013fix}, potentially requiring multiple SPs.

Conversely, CWE-125 (Out-of-bounds Read) and CWE-89 (SQL Injection) are prominent in the general top-10 but absent from the multi-SP top-10. That may be because OOB read bugs tend to be spatially localized~\cite{szekeres2013sok}, and static analysis tools can enumerate the affected buffer access sites, reducing the likelihood of requiring multiple SPs. SQL injection similarly benefits from a well-established fixing pattern (parameterized queries)~\cite{halfond2006classification} that can systematically address injection sites within a single refactoring effort.

\noindent\textbf{SP count distribution.}
\Cref{fig:sp_distribution} shows that 64.4\% of multi-SP CVEs in \spdb have exactly two SPs and only 6.7\% have five or more, indicating that most cases involve few SPs.

\subsubsection{Gaps of Widely Used Vulnerability Databases from \spdb}
\label{sec:db-gaps}
As shown in \Cref{tab:db_merge}, we compare the number of multi-SP CVEs and average SPs per CVE for each database, standalone vs.\ after merging all four sources. Merging reveals patches missed by individual databases, increasing both the number of recognized multi-SP CVEs and the average patch count per multi-SP CVE.

NVD benefits the most (+55.1\% in CVE coverage, from 2.04 to 2.55 avg.\ SPs), while GHSA shows the smallest gain (36.1\%, 17.9\%).

Overall, all four databases exhibit substantial growth after merging, confirming that no single database provides sufficient coverage~\cite{guo2022detecting}, and integrating multiple sources is essential for constructing \spdb.

\mybox{
\textbf{Finding (RQ1):} Multi-SP vulnerabilities are a systemic and intensifying phenomenon: 20.6\% of all CVEs with SPs involve multiple SPs, with annual prevalence rising steadily from 15.9\% in 2020 to 24.5\% in 2024. Yet this phenomenon is severely underreported: merging all four vulnerability databases increases recognized multi-SP CVE counts by 36-55\% over any single source, with NVD showing the largest gain (+55.1\%).
}

\begin{table}[t]
\centering
\setlength{\tabcolsep}{3.5pt}
\caption{Multi-SP CVE coverage and average SPs per CVE for each vulnerability database, before and after merging all four databases.}
\label{tab:db_merge}
\begin{threeparttable}
\begin{tabular}{@{}l rr r rr r@{}}
\toprule
& \multicolumn{3}{c}{\textbf{Multi-SP CVE Coverage}} & \multicolumn{3}{c}{\textbf{Avg.\ SPs per CVE}} \\
\cmidrule(lr){2-4}\cmidrule(lr){5-7}
\textbf{DB} & \textbf{Own\textsuperscript{$\dagger$}} & \textbf{Merged} & \textbf{$\Delta$\%} & \textbf{Own\textsuperscript{$\dagger$}} & \textbf{Merged} & \textbf{$\Delta$\%} \\
\midrule
NVD  & 1,966 & 3,050 & \textbf{+55.1\%} & 2.04 & 2.55 & \textbf{+25.1\%} \\
Snyk & 3,360 & 4,754 & +41.5\% & 2.15 & 2.66 & +23.7\% \\
GHSA & 2,954 & 4,019 & \textbf{+36.1\%} & 2.33 & 2.75 & \textbf{+17.9\%} \\
OSV  & 3,610 & 5,139 & +42.4\% & 2.26 & 2.72 & +20.1\% \\
\bottomrule
\end{tabular}
\begin{tablenotes}\footnotesize
  \item[$\dagger$] Using the database alone, without cross-database merging.
\end{tablenotes}
\end{threeparttable}
\end{table}

\subsection{RQ2: Reasons of Multi-SP Phenomenon}
\label{sec:reasons}

\subsubsection{Taxonomy Construction}
\label{sec:taxonomy_construction}
To systematically understand why a single vulnerability is associated with multiple SPs, we analyzed the phenomenon at the patch-pair level. For each CVE, we ordered its SPs by commit date 
and formed consecutive pairs $(sp_1, sp_2)$. For each pair, we then asked: why is $sp_2$ necessary to remediate the CVE in the presence of $sp_1$? This pair-centric formulation captures the directional dependency between patches and mirrors the practical scenario in which a developer, already aware of an existing fix, still needs to commit an additional one. Based on these patch pairs, we constructed a two-level taxonomy. The details are as follows.

\noindent\textbf{Dataset.}
We conducted a dataset, \sptax, a subset of \spdb, including 4,208 multi-SP CVEs from 2005--2024, for taxonomy construction.
The remaining 2025--2026 CVEs in \spdb are reserved as the sampling pool for the tool evaluation in RQ3, reducing the risk of data leakage.
For a CVE with $n$ SPs, this yields $n-1$ patch pairs, resulting in 6,809 patch pairs in total as the unit of analysis.

We constructed the taxonomy through two steps:

(1) \noindent\textit{Card Sorting.} 
Card sorting~\cite{spencer2009card} provides a bottom-up way to surface pair-specific necessity rationales by obtaining the atomic reasons with LLM assistance (GPT-5.2) at scale and manually consolidating them into a hierarchical taxonomy.
For each patch pair, the LLM is given the full commit context of both $sp_1$ and $sp_2$, including repository metadata, commit messages, and code diffs. It then follows a two-step chain-of-thought (CoT) protocol. First, it analyzes what each patch addresses and derives a pair-specific necessity rationale explaining why $sp_2$ remained necessary after $sp_1$, treating code diffs as primary evidence and commit metadata as supporting evidence. Second, it compares that rationale against the evolving taxonomy and either assigns the pair to an existing sub-category when appropriate or proposes a new one only when the rationale is genuinely distinct from all existing sub-categories. Applying this process sequentially to all patch pairs produced an initial set of 62 candidate sub-categories.
To assess its reliability, two authors with 3+ years of security-analysis experience independently audited a random sample of 50 of the LLM's pair-level outputs against the underlying commit evidence, finding 94\% of them accurate with 96\% raw inter-rater agreement (Cohen's $\kappa$=0.73).

(2) \noindent\textit{Taxonomy Consolidation.} 
These initial sub-categories were intentionally fine-grained to preserve analytical nuance, but they also contained overlaps and redundancies. To organize them into a coherent taxonomy, 
we first introduced a top-level set of (\textbf{categories}) \emph{a priori} to capture the locations of the patch pair at the Git structure, i.e., (1) within a single branch/tag of one repository; (2) parallel patching across multiple branches of the same repository; (3) parallel patching across multiple repositories.
During inspection of the initial sub-categories, we also identified three kinds of cases that were not adequately captured by Git structure alone: \textit{duplicate fixes within the same branch}, \textit{non-security fixes}, and \textit{others}. We therefore added them as supplementary categories to ensure full coverage.

After partitioning patch pairs by top-level category, we consolidated the fine-grained categories within each partition into a set of \textbf{sub-categories}, following an established taxonomy development method~\cite{nickerson2013method}.
At this level, each sub-category captures the specific reason why an additional SP was needed within the given patching scope.
Two authors independently performed the consolidation by merging, splitting, renaming, or discarding candidate sub-categories based on semantic distinctness, using the guiding criterion that no patch pair should simultaneously satisfy the definitions of two different sub-categories. We iterated until theoretical saturation, i.e., additional patch-pair evidence produced no distinct sub-categories~\cite{glaser2017discovery}, achieving a Cohen's $\kappa$ of 0.88~\cite{landis1977measurement}; remaining disagreements were adjudicated by a third expert, yielding the final taxonomy of 6 categories and 16 sub-categories.

\subsubsection{Taxonomy Overview}

\begin{figure}[t]
    \centering
    \includegraphics[width=\columnwidth]{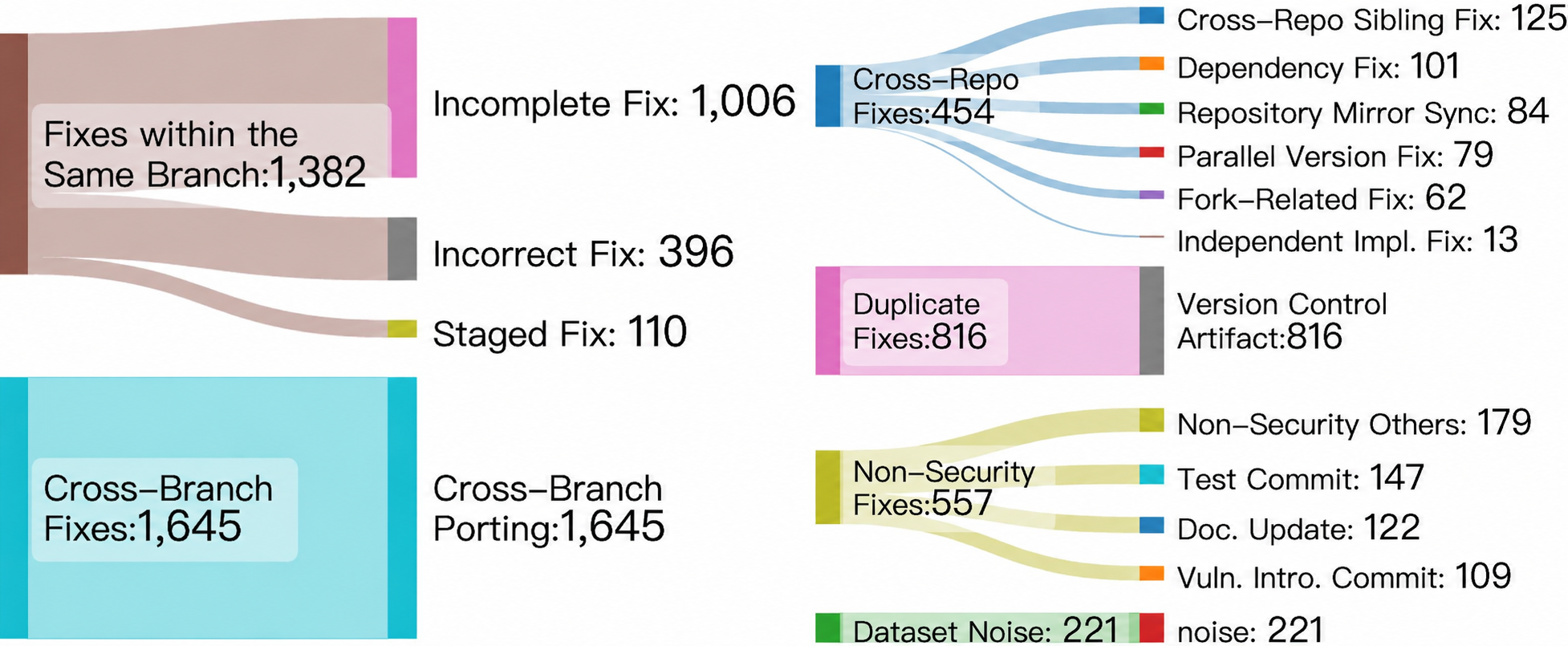}
    \caption{Taxonomy of multi-SP reasons (\#CVEs).}
    \label{fig:sankey_taxonomy}
\end{figure}

The final taxonomy comprises 6 categories and 16 sub-categories (\Cref{fig:sankey_taxonomy}).
At the top level, the category captures the roughly coarse-grained scopes of patch pairs. Within each category, the sub-category captures the specific rationale for why an additional SP was still necessary within that scope.

\noindent\textbf{C1: Fixes within the Same Branch.}
This category covers cases in which multiple SPs for the same CVE occur in the same repository and on the same branch (or tag). Unlike cross-branch or cross-repository cases, the additional patch is required within a single fixing context, meaning that the need for $sp_2$ arises from how the vulnerability is fixed in that codebase. We identified 3 sub-categories that captured the specific rationale for why $sp_2$ remained necessary after $sp_1$, including incomplete, incorrect, and staged fixes.

\noindent $\bullet$ \textbf{Incomplete Fix} (1,006 CVEs, 1,396 patch pairs).
\textit{Incomplete Fix} is the dominant same-branch pattern. In this case, $sp_1$ adopts an appropriate fixing strategy, but leaves part of the attack surface exposed by missing vulnerable locations, execution paths, edge cases, or input variants. This category is especially important because it reflects a \textit{true patching gap}: users and downstream tools may treat $sp_1$ as the fix even though exploitable behavior remains.

A representative example is CVE-2020-10235 in Froxlor (Figure~\ref{fig:incomplete-fix-example}), where the initial patch sanitized only two of three user-controlled arguments in a shell command, leaving the password parameter injectable. A follow-up commit was needed to apply the same \texttt{escapeshellarg()} call to the overlooked variable.

\input{imgs/incomplete_fix_example.tex}

\noindent $\bullet$ \textbf{Incorrect Fix} (396 CVEs, 419 patch pairs).
\textit{Incorrect Fix} captures a different failure mode: $sp_1$ attempts to remediate the vulnerability, but its implementation is itself flawed, for example, due to a logic error, typo, overly restrictive or overly permissive check, functional regression, or a misidentified root cause. Accordingly, $sp_2$ does not merely extend the protection introduced by $sp_1$; instead, it \textit{modifies, reverts, or replaces} the faulty security logic.

\noindent $\bullet$ \textbf{Staged Fix} (110 CVEs, 134 patch pairs).
\textit{Staged Fix} covers cases where the patching is decomposed into structurally dependent steps. Here, $sp_1$ lays the groundwork and $sp_2$ completes the fix on top of it; the latter could not be implemented without the former. 
Together, the two commits form a single logical security fix that is deliberately split across commits for implementation reasons.

\noindent\textbf{C2: Cross-Branch Fixes.}
It covers cases in which the same CVE is remediated across multiple maintained branches of a single repository. Such cases arise naturally in projects that simultaneously maintain LTS, stable, and legacy release lines~\cite{tan2022understanding,yang2023enhancing,liu2025patchscope}.

\noindent $\bullet$ \textbf{Cross-Branch Porting} (1,645 CVEs, 2,357 patch pairs).
It captures the propagation of an established fix from one branch to another maintained branch in the same repository. Here, the necessity of $sp_2$ stems from the need to apply the same fix to another supported development line. Depending on branch divergence, $sp_2$ may appear either as a near-identical cherry-pick or as a substantially adapted re-implementation~\cite{pan2024automating,wu2025mystique}.

An example is CVE-2019-20920 in Handlebars.js (Figure~\ref{fig:cross-branch-example}). On \texttt{4.x}, the fix was a one-line \texttt{String()} coercion in the \texttt{lookup} helper closing a \texttt{toString()} bypass of the constructor check.
On \texttt{3.x}, however, the same vulnerability required a far more extensive patch: blocking hazardous properties and adding \texttt{hasOwnProperty} guards in both the runtime and the compiler.
This illustrates that the same vulnerability can manifest as substantially different patches depending on the security maturity of each branch.

\input{imgs/cross_branch_example.tex}

\noindent\textbf{C3: Cross-Repository Fixes.}
This category covers cases in which a single vulnerability affects multiple repositories.
Unlike cross-branch fixes (C2), C3 concerns vulnerabilities whose patching footprint spans distinct repositories.

\noindent $\bullet$ \textbf{Sibling/Component Fix} (125 CVEs, 167 patch pairs).
As the most common cross-repository pattern, it captures vulnerabilities that manifest across multiple components, plug-ins, or packages in the same ecosystem, each maintained in its own repository.

\noindent $\bullet$ \textbf{Dependency Fix} (101 CVEs, 106 patch pairs).
It captures patches that propagate along an upstream-downstream dependency relationship. $sp_1$ fixes the vulnerability in an upstream library, while $sp_2$ brings that patch into a downstream project through a version bump~\cite{zhang2023mitigating,zhang2023compatible}, vendored code update, or local port.

\noindent $\bullet$ \textbf{Repository Mirror Sync} (84 CVEs, 101 patch pairs).
It refers to cases where substantially the same codebase is distributed across multiple repositories due to mirroring, monorepo splitting, or read-only sub-repository extraction.

\noindent $\bullet$ \textbf{Parallel Version Fix} (79 CVEs, 120 patch pairs).
It captures projects that maintain different major versions in separate repositories, analogous to cross-branch maintenance, except that the version lines are separated at the repository level.

\noindent $\bullet$ \textbf{Fork-Related Fix} (62 CVEs, 75 patch pairs).
It covers cases where one affected repository is a fork, or otherwise derived from, the other, by the explicit fork relationship.

\noindent $\bullet$ \textbf{Independent Implementation Fix} (13 CVEs, 13 patch pairs).
It covers independent projects that implement the same protocol or specification and thus exhibit the same vulnerability despite no code sharing, fork lineage, or dependency relation.

\noindent\textbf{C4: Duplicate Fixes within the Same Branch.}
This category captures cases where two SP entries on the same branch correspond to the same underlying fix, with one sometimes subsuming the other. 
A sub-category is \textit{Version Control Artifact} (816 CVEs, 1,070 Patch Pairs), where $sp_1$ and $sp_2$ are duplicate representations of the same security fix, typically produced by version-control workflows such as \texttt{merge}, \texttt{rebase}, or \texttt{squash}. Although such workflows may span multiple branch histories, they are treated as the same-branch duplicates because their purpose is to update a single release line rather than to introduce a separate fix for another branch or repository.

\noindent\textbf{C5: Non-Security Fixes.}
These are patch pairs in which at least one linked commit was not itself a security fix, but collateral work surrounding the actual fix.
We observed four sub-categories: \textit{Test Commit}, \textit{Vulnerability-Introducing Commit}, \textit{Documentation/Advisory Update}, and \textit{Non-Security Others}, which mainly cover tests, bug-introducing commits, advisory-only changes, and post-fix engineering follow-up. This suggests that CVE-related references often mix the actual fix with surrounding non-security activity.

\noindent\textbf{C6: Dataset Noise} (221 CVEs, 235 patch pairs).
Unlike C1--C5, which require a reliable intentional or causal relation between the patch pair and the target CVE, C6 captures dataset-noise pairs where at least one commit lacks such a relation, such as unrelated commits, security patches for other CVEs, or commits accidentally linked through shared PRs or releases.

\noindent\textbf{Heterogeneity across Languages and Ecosystems.}
The prevalence of each category shifts systematically with implementation language and project ecosystem. Cross-branch fixes occur in 57.5\% of Python CVEs, against 39.09\% across all CVEs, whereas cross-repository fixes occur in only 5.2\% (overall 10.79\%), a profile of single-repository projects that port across several maintained branches. C and C++ invert this pattern: fixes within the same branch occur in 43.6\% and 44.2\% of their CVEs (overall 32.84\%), and C's cross-repository share of 20.2\% is nearly twice the overall rate, consistent with an ecosystem that lacks a unified package manager and propagates code through vendoring and forks. At the ecosystem level, nearly every vulnerability in projects that maintain several release lines in parallel requires a cross-branch fix (98.2\% of Django CVEs, 88.2\% of Tomcat CVEs, overall 39.09\%). Among cross-repository fixes, the majority (57.3\%) involve repositories under the same organization (release-line mirrors, package splits, and plugin-host pairs).

\subsubsection{Reliability Validation}
\label{sec:reliability_validation}
To validate the clarity and consistent applicability of our two-level taxonomy, two authors independently classified a stratified random sample of 200 patch pairs using only the sub-category definitions, achieving a Cohen's $\kappa$ of 0.95 on the 16 sub-categories. A third author adjudicated the disagreements, yielding a final classification precision of 96.0\% (192/200).

\mybox{\textbf{Finding (RQ2):} 
We construct a two-level taxonomy of the reasons behind the multi-SP phenomenon, including 6 categories and 16 sub-categories.
\textit{Cross-branch fixes} are the dominant reason (1,645 CVEs, 39.09\%).
\textit{Fixes within the same branch} are the second most common (1,382 CVEs, 32.84\%), with \textit{Incomplete Fix} and \textit{Incorrect Fix} being the most prevalent subtypes. \textit{Cross-repository fixes} affect a further 454 CVEs.
Besides, a nontrivial portion is attributable to version-control artifacts that duplicate fixes and to non-security commits.
}

\subsection{RQ3: Finding-Driven Prototype Development and Evaluation}

\begin{figure*}
  \centering
  \includegraphics[width=0.9\textwidth]{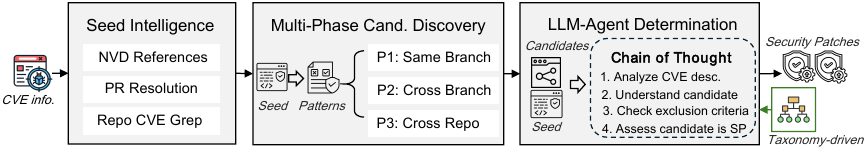}
  \caption{Overview of \tool: a taxonomy-driven prototype for security patch discovery given a CVE ID.}
  \label{fig:tool_overview}
\end{figure*}

\subsubsection{Design of~\tool}
\label{sec:approach}
We design \tool as a fully automated end-to-end pipeline that, given only a CVE-ID, discovers a comprehensive set of security patches.
\tool focuses exclusively on \textit{code-level} security fixes, i.e., commits whose diffs modify source files. Commits touching only documentation, tests, build configurations, or dependency lock files are excluded, as they do not constitute actionable security patches for downstream consumers.

As shown in~\Cref{fig:tool_overview}, the pipeline consists of three main stages: Stage I for \textit{Seed Intelligence} (discovering initial seed patch commits from public sources) and Stage II for \textit{Multi-Phase Candidate Discovery} (systematically finding candidate commits), followed by Stage III for \textit{LLM-Agent Determination} (determining whether each candidate commit is a true code-level security patch for the target vulnerability).
Crucially, the entire pipeline is \textit{taxonomy-driven}: the two-level reason taxonomy established in Section~\ref{sec:taxonomy_construction}, primarily its three categories (C1: fixes within the same branch, C2: cross-branch fixes, and C3: cross-repository fixes).

\paragraph{Stage I: Seed Intelligence.}
Given only a CVE-ID, \tool automatically discovers seed patch commits (hereafter, \textit{seeds}) by querying multiple public intelligence sources in sequence:
\begin{enumerate}[leftmargin=*,nosep,label=\textbf{S\arabic*}.]
  \item \textbf{Vulnerability Database References}: Extract commit URLs from the reference lists of public vulnerability databases (e.g., NVD) and security advisory platforms.
  \item \textbf{Pull Request Resolution}: For references that link to pull requests rather than commits, resolve the PR via the GitHub API to obtain merge commits and individual PR commits.
  \item \textbf{Repository CVE Grep}: For repositories identified through S1--S2, search the local git clone for commits mentioning the CVE-ID in their message (\texttt{git log --grep}).
\end{enumerate}

To ensure that seeds correspond to actionable code-level SPs, we filter out commits whose changed files are entirely non-code, such as documentation, test-only changes,lock files, or empty merge commits, removing noise from advisory references that point to changelog updates or version bumps rather than actual code fixes.

\paragraph{Stage II: Multi-Phase Candidate Discovery.}
\label{sec:candidate-discovery}
\tool applies three discovery phases per seed and merges the resulting candidate commits.

\noindent\textbf{Phase~1: Same-Branch Candidate Discovery} targets \textit{C1 (fixes within the same branch)} scenarios.
Within a configurable time window ($\pm$365 days, with a maximum of 2,000 commits) around each seed on the same branch, \tool searches for candidate commits. Compared with Prospector's $\pm$60-day search window~\cite{sabetta2024known}, we adopt a more conservative $\pm$365-day window because over 95\% of multi-SP CVEs in \spdb have their first and last recorded SP within one year, and cap the scan at 2,000 commits to bound the search cost in highly active projects. \tool uses the following candidate-discovery patterns:
\begin{itemize}[leftmargin=*,nosep]
  \item \textit{C1-1: Time proximity with multi-signal support}. Commits that are either in a direct parent-child relationship with the seed, or occur within 7 days of the seed and share at least one corroborating signal, including the same author, overlapping code files, or overlapping modified hunks.
  \item \textit{C1-2: Identical commit message}. Commits whose normalized message is identical to that of the seed.
  \item \textit{C1-3/C1-6: CVE-ID/Issue-ID grep}. Commits mentioning the same CVE-ID or issue tracker ID in their message.
  \item \textit{C1-4: Code evolution}. Commits that last modified the same lines changed by the seed, identified through \texttt{git blame} tracing.
  \item \textit{C1-5: Same pull request}. Commits associated with the same PR as the seed, identified via the GitHub API.
  \item \textit{C1-7: Commit hash reference}. Commits that reference the seed's commit hash in their message, indicating an explicit relationship.
  \item \textit{C1-8: Revert Detection}. Commits that revert the seed, which may indicate that the initial fix was later replaced or corrected.
  \item \textit{C1-9: Fuzzy commit message with multi-signal support}. Commits with a normalized message similar to that of the seed (threshold = 0.65) and at least one additional supporting signal, such as the same author or overlapping code files.
\end{itemize}

\noindent\textbf{Phase~2: Cross-Branch Candidate Discovery} targets \textit{C2 (cross-branch fixes)} scenarios.
\tool selects branches whose release versions differ from the seed branch, and collects up to 2,000 commits modifying files with the same basenames as those modified by the seed commit, prioritized by temporal proximity to the seed commit. Within this set, we search for candidates using the following candidate-discovery patterns:
\begin{itemize}[leftmargin=*,nosep]
  \item \textit{C2-1: Fuzzy commit message with multi-signal support}. Same as \textit{C1-9}.
  \item \textit{C2-2: Cherry-pick detection}. Commits reachable from the seed via BFS-based cherry-pick transitive closure, using \texttt{cherry picked from commit} trailers and reverse hash references.
  \item \textit{C2-3: Overlapping code changes}. Commits whose code hunks are similar with those of the seed, indicating semantically equivalent changes applied independently to separate branches.
  \item \textit{C2-4/C2-5: Issue-ID/CVE-ID grep}. Same as \textit{C1-6/C1-3}.
\end{itemize}

\noindent\textbf{Phase~3: Cross-Repository Candidate Discovery} targets \textit{C3 (cross-repository fixes)} scenarios.
Our analysis in Section~\ref{sec:reasons} shows that the six sub-categories of C3 in \sptax are not distributed arbitrarily across repositories; instead, they typically occur between repositories with some relationships, such as parallel-version repositories or synchronized repositories.

Motivated by this observation, we maintain a repository-relation table that records repository pairs previously observed in \sptax.
Given the repository containing the seed, \tool first queries this table to find potential repositories.
If no match is found, \tool falls back to the top-5 starred fork-related repositories returned by the GitHub API.
Then we search for candidates within the target repositories using the following candidate-discovery patterns: \textit{C3-1: Identical commit message},\textit{C3-2: Overlapping code changes}, \textit{C3-3/C3-6: CVE-ID/Issue-ID grep}, \textit{C3-4: Cherry-pick detection}, and \textit{C3-5: Identical author date}. \tool considers commits whose author-date timestamps exactly match that of the seed's as candidates, since commits derived from the same original fix across repositories often retain it.

Overall, this stage uses candidate-discovery patterns to reduce the raw commit space into a compact candidate pool for subsequent determination. The patterns rely on observable evidence from local Git histories and GitHub metadata, such as repository history, pull-request metadata, commit messages, and local diff/code search, so candidate discovery itself incurs no LLM cost.

\paragraph{Stage III: LLM-Agent Determination.}
\label{sec:llm-agent-determination}
Each candidate identified by the discovery phases above is determined by an LLM-based agent before being added to the output.
The agent operates with \textit{taxonomy-informed prompts}: depending on whether the candidate resides on the same branch, a different branch, or a different repository as the seed, one of three specialized system prompts is selected.
Each prompt combines exclusion criteria with taxonomy-derived acceptance criteria:  a candidate is retained only if it is not excluded and is supported by one sub-category.

The exclusion criteria remove candidates that should not be linked to the input CVE: (1) commits that are not security patches at all, such as tests, documentation, cleanup, version bumps, logging, refactoring, build/CI changes, or vulnerability-introducing commits; and (2) security-related commits that are not clearly tied to the input CVE. For the latter, the agent is explicitly told that temporal proximity, shared authorship, overlapping files, similar paths, or a similar bug class alone are insufficient evidence of the same CVE.

The acceptance criteria are derived from the sub-categories of the corresponding taxonomy partition. For example, a same-branch candidate is accepted only when the evidence linking it to the seed patch supports sub-categories such as incomplete fix, incorrect fix, or staged fix. Thus, a candidate is accepted only if no exclusion criterion applies and there is positive evidence that its relation to the seed patch satisfies one of the taxonomy-derived acceptance criteria for the corresponding scenario.

The agent follows a four-step CoT protocol: (i)~analyze the vulnerability from the CVE description, (ii)~examine what the candidate commit addresses, (iii)~check exclusion criteria, and (iv)~match the candidate against the taxonomy-driven acceptance criteria.
The agent returns a structured JSON verdict containing the classification (\textit{yes}, \textit{no}, or \textit{uncertain}), and key evidence excerpts from code diffs or commit messages.
Only when the evidence is genuinely insufficient (e.g., empty or truncated diffs) does the agent return \textit{uncertain}.
Only candidates receiving a \textit{yes} verdict are retained; \textit{no} and \textit{uncertain} verdicts are not included in the output.

\subsubsection{Effectiveness Evaluation of~\tool}
\label{sec:rq3_eval}
We evaluate \tool as a standalone end-to-end pipeline on a dataset with other baselines: given only a CVE-ID, the tool autonomously discovers seed patch commits from public sources, and expands them into a comprehensive set of related security patches.

\noindent\textbf{Dataset.}
We construct \speval by randomly sampling 300 multi-SP CVEs from the late-2025--2026 portion of \spdb, which keeps it temporally disjoint from \sptax (2005--2024) to avoid data leakage while remaining feasible for manual verification. 
\speval (2 SPs: 58.7\%, 3+ SPs: 41.3\%) is broadly consistent with the patch-count distribution of \spdb shown in Figure~\ref{fig:sp_distribution}.
We then partition the sampled CVEs by their ground-truth patch locations:
(1)~\textit{Same-repository} (239 CVEs), where all ground-truth SPs reside in a single GitHub repository and cover same-branch or cross-branch scenarios;
(2)~\textit{Cross-repository} (61 CVEs), where SPs span two or more GitHub repositories.
After removing same-branch duplicates and test-/documentation-only commits, we manually validated the remaining 800 SP candidates to filter non-security commits (e.g., vulnerability-introducing commits) and dataset noise defined in our taxonomy (\Cref{sec:reasons}). Two authors independently reviewed the candidates (Cohen's $\kappa$ = 0.945), with disagreements resolved by a third author, yielding 719 ground-truth SPs for 300 CVEs.

\noindent\textbf{Baselines.}
We selected baselines from the line of work most directly related to ours, i.e., tracing security patches for disclosed vulnerabilities (Section~\ref{sec:related_localization}), covering representative reproducible families: Prospector~\cite{sabetta2024known} (heuristic/rule-based tracing), PatchFinder~\cite{li2024patchfinder} (LLM-based CVE--commit matching), and SHIP~\cite{song2025not} (multi-SP-oriented localization). We excluded other recent methods that either lack a public implementation (e.g., PromVPat~\cite{zhang2024dual}, Taper~\cite{ran2025efficient}, SPV~\cite{wang2025locating}) or whose released artifact we could not reproduce during our replication attempt (e.g., PatchSeeker~\cite{nguyen2025patchseeker}). The details are as follows.
\begin{itemize}[leftmargin=*,nosep]
  \item \textbf{PatchFinder}~\cite{li2024patchfinder}: ranks candidate commits using a fine-tuned LLM that scores commits by similarity to CVE descriptions.
  \item \textbf{SHIP}~\cite{song2025not}: an LLM-enhanced patch localization approach that explicitly targets multi-SP vulnerabilities, producing a ranked list of groups of candidate commits per repository.
  \item \textbf{Prospector}~\cite{sabetta2024known}: applies rule-based heuristics (e.g., advisory references, message keywords) to rank candidates.
  \item \textbf{LLM-Only}: an ablation using the same seed intelligence and candidate pool as \tool, but replacing candidate-discovery patterns and determination with a single GPT-5-mini judgment for each candidate based on the CVE description, seed patch, and candidate commit. This isolates the benefit of \tool's pattern-based filtering and taxonomy-guided reasoning. Due to its high cost (\textasciitilde108K API calls for 100 CVEs, 67.5$\times$ more than \tool), we evaluate it on a 100-CVE subset (80 same-repo + 20 cross-repo).
  \item \textbf{NVD seeds (floor)}: establishes the baseline achievable by simply querying NVD references, without any expansion or ranking.
\end{itemize}

The first three baselines take a CVE-ID and repository URL as input, producing a ranked candidate list, while LLM-Only produces the output with the same format as \tool.

\noindent\textbf{Metrics.}
We report \textit{recall} (fraction of ground-truth SPs found) and \textit{precision} (fraction of outputs that are true positives).
Baselines produce ranked lists; we evaluate at $K{=}10$ and $K{=}50$.
\tool produces an unordered set capped at 10 (avg.\ 3.6 patches/CVE).
We report the NVD seeds (floor): 
the recall achievable by simply retrieving commit URLs from NVD references.

\noindent\textbf{Environment.}
All experiments run on an Ubuntu 24.04 server with 128 CPU cores, 256\,GB RAM, and 8 RTX A6000 GPUs. We use the released implementations and default configurations of PatchFinder~\cite{li2024patchfinder} and Prospector~\cite{sabetta2024known}. 
As SHIP~\cite{song2025not} provides no public pre-trained model, we retrain it with the authors' released scripts and reported configuration (8:1:1 split, lr 1e-4, 20 iterations, batch 48), without tuning on our dataset. 
\tool uses 20 parallel workers with a 360-second timeout per CVE; LLM determination (Section~\ref{sec:llm-agent-determination}) uses \texttt{gpt-5-mini-2025-08-07}~\cite{openrouter2025gpt5mini,openai2025gpt5developers} with temperature~0.

\noindent\textbf{Manual Verification of Ground Truth.}
A core premise of our study is that even when existing vulnerability databases do provide SPs for a vulnerability, those patches may still be incomplete.
Consequently, \speval's initial ground truth $GT$ cannot be assumed to be complete either: patches that any evaluated baselines discovers but $GT$ does not contain are not necessarily false positives; they may be genuine not-recorded SPs.
We therefore adopt a two-phase protocol: 
we first evaluate all tools against the original $GT$ on equal footing denoted as \texttt{Before manual}.
We then perform a balanced ground-truth expansion over the full 300-CVE evaluation set. For \tool, we inspect 516 outputs that passed LLM determination but were absent from $GT$. For each ranking-based baseline (Prospector, PatchFinder, and SHIP), we inspect the top-ranked output for each evaluation query, resulting in 358 candidates from Prospector, 374 from PatchFinder, and 374 from SHIP.
Specifically, 315 out-of-$GT$ candidates produced by \tool are confirmed as SPs. The inspected outputs of Prospector, PatchFinder, and SHIP contain 202, 28, and 156 confirmed SPs, respectively, although most of these baseline positives are already present in the original $GT$. After merging duplicate confirmations across tools, this manual expansion identifies 335 previously unrecorded SPs, increasing the ground truth to 1,054 SPs in $GT'$, against which all tools are re-evaluated under the \texttt{After manual} setting.

\begin{table}[t]
\centering
\caption{Effectiveness on \speval (300 multi-SP CVEs) with ground truth before (abbr. GT) and after (abbr. GT') manual verification, comparing \tool against three baselines (shown at $K{=}10$ and $K{=}50$).}
\label{tab:rq3_tools}
\begin{threeparttable}
\setlength{\tabcolsep}{3.5pt}
\begin{tabular}{@{}l l l rr rr@{}}
\toprule
& & & \multicolumn{2}{c}{\textbf{Same-repo (239)}} & \multicolumn{2}{c}{\textbf{Cross-repo (61)}} \\
\cmidrule(lr){4-5}\cmidrule(lr){6-7}
& \textbf{Tool} & $K$ & \textbf{Recall} & \textbf{Prec.} & \textbf{Recall} & \textbf{Prec.} \\
\midrule
\multirow{9}{*}{\rotatebox[origin=c]{90}{\scriptsize\textit{Before ($GT$, 719)}}}
& NVD seeds (floor)  & ---  & 0.149 & --- & 0.179 & --- \\
\cmidrule{2-7}
& PatchFinder       & 10  & 0.138 & 0.034 & 0.057\textsuperscript{$\dagger$} & 0.013\textsuperscript{$\dagger$} \\
& SHIP              & 10  & 0.509 & 0.125 & 0.400\textsuperscript{$\dagger$} & 0.093\textsuperscript{$\dagger$} \\
& Prospector        & 10  & 0.411 & 0.101 & 0.543\textsuperscript{$\dagger$} & 0.127\textsuperscript{$\dagger$} \\
\cmidrule{2-7}
& PatchFinder       & 50  & 0.363 & 0.018 & 0.236\textsuperscript{$\dagger$} & 0.011\textsuperscript{$\dagger$} \\
& SHIP              & 50  & 0.535 & 0.026 & 0.429\textsuperscript{$\dagger$} & 0.020\textsuperscript{$\dagger$} \\
& Prospector        & 50  & 0.504 & 0.025 & 0.650\textsuperscript{$\dagger$} & 0.031\textsuperscript{$\dagger$} \\
\cmidrule{2-7}
& \tool             & --- & \textbf{0.910} & \textbf{0.510} & \textbf{0.829} & \textbf{0.298} \\
\midrule\midrule
\multirow{9}{*}{\rotatebox[origin=c]{90}{\scriptsize\textit{After ($GT'$, 1054)}}}
& NVD seeds (floor)  & ---  & 0.107\textsubscript{\textcolor{red}{$\downarrow$}} & --- & 0.119\textsubscript{\textcolor{red}{$\downarrow$}} & --- \\
\cmidrule{2-7}
& PatchFinder       & 10  & 0.115\textsubscript{\textcolor{red}{$\downarrow$}} & 0.039 & 0.061\textsuperscript{$\dagger$} & 0.025\textsuperscript{$\dagger$} \\
& SHIP              & 10  & 0.391\textsubscript{\textcolor{red}{$\downarrow$}} & 0.134 & 0.332\textsubscript{\textcolor{red}{$\downarrow$}}\textsuperscript{$\dagger$} & 0.135\textsuperscript{$\dagger$} \\
& Prospector        & 10  & 0.310\textsubscript{\textcolor{red}{$\downarrow$}} & 0.106 & 0.385\textsubscript{\textcolor{red}{$\downarrow$}}\textsuperscript{$\dagger$} & 0.157\textsuperscript{$\dagger$} \\
\cmidrule{2-7}
& PatchFinder       & 50  & 0.289\textsubscript{\textcolor{red}{$\downarrow$}} & 0.020 & 0.193\textsubscript{\textcolor{red}{$\downarrow$}}\textsuperscript{$\dagger$} & 0.016\textsuperscript{$\dagger$} \\
& SHIP              & 50  & 0.416\textsubscript{\textcolor{red}{$\downarrow$}} & 0.028 & 0.369\textsubscript{\textcolor{red}{$\downarrow$}}\textsuperscript{$\dagger$} & 0.030\textsuperscript{$\dagger$} \\
& Prospector        & 50  & 0.386\textsubscript{\textcolor{red}{$\downarrow$}} & 0.026 & 0.475\textsubscript{\textcolor{red}{$\downarrow$}}\textsuperscript{$\dagger$} & 0.039\textsuperscript{$\dagger$} \\
\cmidrule{2-7}
& \tool             & --- & \textbf{0.927}\textsubscript{\textcolor{green!60!black}{$\uparrow$}} & \textbf{0.725}\textsubscript{\textcolor{green!60!black}{$\uparrow$}} & \textbf{0.873}\textsubscript{\textcolor{green!60!black}{$\uparrow$}} & \textbf{0.548}\textsubscript{\textcolor{green!60!black}{$\uparrow$}} \\
\bottomrule
\end{tabular}
\begin{tablenotes}\footnotesize
  \item[$\dagger$] Baselines are single-repository tools; for cross-repo CVEs, we run them on each NVD-referenced repository and interleave the results round-robin, so they cannot discover SPs in repositories not already listed by NVD.
\end{tablenotes}
\end{threeparttable}
\end{table}

\begin{table}[t]
\centering
\caption{Cost-effectiveness of \tool versus the LLM-Only baseline on a random 100-CVE subset of \speval, with GT and GT'.}
\label{tab:rq3_llmonly}
\begin{threeparttable}
\setlength{\tabcolsep}{3.5pt}
\begin{tabular}{@{}l l rr rr@{}}
\toprule
& & \multicolumn{2}{c}{\textbf{Same-repo (80)}} & \multicolumn{2}{c}{\textbf{Cross-repo (20)}} \\
\cmidrule(lr){3-4}\cmidrule(lr){5-6}
& \textbf{Tool} & \textbf{Recall} & \textbf{Prec.} & \textbf{Recall} & \textbf{Prec.} \\
\midrule
\multirow{2}{*}{\rotatebox[origin=c]{90}{\scriptsize $GT$}}
& LLM-Only & 0.821 & 0.491 & 0.771 & 0.180 \\
& \tool  & \textbf{0.937} & 0.488 & \textbf{0.812} & \textbf{0.228} \\
\cmidrule{2-6}
\multirow{2}{*}{\rotatebox[origin=c]{90}{\scriptsize $GT'$}}
& LLM-Only & 0.688 & 0.604 & 0.663 & 0.286 \\
& \tool  & \textbf{0.950} & \textbf{0.726} & \textbf{0.888} & \textbf{0.462} \\
\bottomrule
\end{tabular}
\end{threeparttable}
\end{table}

\paragraph{Results Before Manual Verification}
\label{sec:rq3_before}

\Cref{tab:rq3_tools} (upper section) shows each tool's output evaluated against $GT$.
On same-repo CVEs, \tool (0.910 recall) discovers more ground-truth SPs than the representative ranking-based baselines, including SHIP at $K{=}50$ (0.535 recall), while maintaining 0.510 precision.
Increasing $K$ from 10 to 50 improves baselines' recall only marginally while causing their precision to collapse below 4\%, indicating that the additional candidates are overwhelmingly noise.
On cross-repo CVEs, \tool achieves 0.829 recall, while Prospector at $K{=}50$ (0.650 recall) with far higher precision.
Note that these baselines are designed for single-repository patch discovery. For CVEs with cross-repository scenarios, we ran them separately on each NVD-referenced repository and interleaved the results; they still cannot discover SPs in repositories not already listed by NVD.

\paragraph{Manual Verification Results}
\label{sec:rq3_manual}
Following the manual verification protocol described above, we identified \textbf{335 previously unrecorded security patches} in \speval, including 315 SPs discovered by \tool.
As shown in \Cref{tab:rq3_missing}, most missing SPs are \textit{cross-repository fixes} (206, 61.5\%), followed by \textit{fixes within the same branch} (85, 25.4\%) and \textit{cross-branch fixes} (44, 13.1\%).
Among all patterns, C3-1 (\textit{Cross-repo (fuzzy message)}) is the largest, accounting for 131 missing SPs.
These results show that many of \tool's apparent false positives against $GT$ are in fact genuine SPs missed by existing databases.

\begin{table}[t]
\centering
\caption{Distribution of 335 ground-truth-missing patches in \speval, grouped by candidate-discovery patterns. N/A denotes the SPs contributed only by the baselines, which are not captured by patterns.}
\label{tab:rq3_missing}
\begin{tabular}{llr}
\toprule
\textbf{Pattern} & \textbf{Description} & \textbf{Count} \\
\midrule
C1-1 & Same-branch (time proximity)    & 35 \\
C1-4 & Same-branch (code evolution)    & 17 \\
C1-6 & Same-branch (CVE-ID grep)       & 10 \\
C1-2 & Same-branch (identical message) & 9  \\
C1-3 & Same-branch (issue-ID grep)     & 7  \\
C1-5 & Same-branch (same PR)           & 2  \\
N/A  & Same-branch (baseline-only)     & 5  \\
\midrule
C2-3 & Cross-branch (overlapping code) & 24 \\
C2-1 & Cross-branch (fuzzy message)    & 15 \\
N/A  & Cross-branch (baseline-only)    & 5  \\
\midrule
C3-1 & Cross-repo (fuzzy message)      & 131 \\
C3-2 & Cross-repo (cherry-pick)        & 28 \\
C3-5 & Cross-repo (identical author date) & 26 \\
C3-4 & Cross-repo (issue-ID grep)      & 7  \\
C3-3 & Cross-repo (CVE-ID grep)        & 4  \\
N/A  & Cross-repo (baseline-only)      & 10 \\
\midrule
\multicolumn{2}{l}{\textbf{Total}} & \textbf{335} \\
\bottomrule
\end{tabular}
\end{table}

Based on $GT'$, we re-evaluate all tools without changing their outputs (\Cref{tab:rq3_tools}, lower section). The key observations are:
\begin{itemize}[leftmargin=*,nosep]
  \item \textbf{GT correction validates \tool's output quality.}
  $GT'$ is built by verifying all four tools' outputs, yet the baselines contribute only 20 of its 335 additions, so their recall descreases as $GT'$ grows beyond what their unchanged outputs cover.
  \tool is the only tool whose recall and precision both increase (0.910$\to$0.927 recall, 0.510$\to$0.725 precision on same-repo), confirming that its apparent false positives against $GT$ were in fact genuine SPs missed by databases and other tools.
  \item \textbf{Higher recall with far fewer candidates.} Against $GT'$, \tool (0.927 same-repo recall, 0.873 cross-repo recall) provides higher multi-SP coverage than every baseline even at $K{=}50$, while outputting far fewer candidates. This confirms that candidate-discovery patterns (Section~\ref{sec:candidate-discovery}) are more effective than generic commit ranking for multi-SP discovery. However, \tool still misses 90 SPs in $GT'$. These FNs are mainly same-repository fixes (66\%); the remaining cases (34\%) are cross-repository fixes, mostly within the same organization. The dominant cause is candidate-discovery failure: 92\% of missed patches never enter the candidate pool, often because the seed patch does not provide enough observable evidence to connect related candidates, such as generic commit messages, overlapping code changes, or explicit repository relations. Only 8\% of the missed patches are generated as candidates but rejected during LLM determination.
  \item \textbf{Substantial lift over NVD seeds (floor).} Against $GT'$, \tool achieves +0.820 same-repo recall and +0.694 cross-repo recall over the NVD seeds (floor), demonstrating that the multi-phase candidate discovery phases search for patches far beyond what public database references provide.

  \item \textbf{LLM-only determination is costly with marginal benefit.} The LLM-Only baseline (\Cref{tab:rq3_llmonly}) uses \tool's seed intelligence and candidate pool as \tool, but omits the pattern-based filtering and the taxonomy-driven determination.
  On the same 100-CVE subset, \tool achieves both higher performance for same-repo and cross-repo scenarios, while its candidate-discovery patterns substantially prune the raw search space (up to 2,000 commits per search scope) before LLM determination, resulting in 67.5$\times$ fewer LLM calls (\textasciitilde1.6K vs.\ \textasciitilde108K). This demonstrates that domain-specific candidate-discovery patterns are both more effective and more efficient than brute-force LLM determination.
\end{itemize}

\mybox{
\textbf{Finding (RQ3):} On 300 multi-SP CVEs with manually verified ground-truth SPs, \tool achieves 0.927 same-repo recall and 0.873 cross-repo recall, improving multi-SP patch coverage over all baselines even at $K{=}50$.
Compared with an LLM-only ablation, \tool achieves higher recall and precision while using 67.5$\times$ fewer LLM calls, confirming that candidate-discovery patterns are both more effective and more cost-efficient than brute-force LLM determination.
Manual verification confirms \textbf{335 previously unrecorded SPs} absent from all vulnerability databases, including 315 contributed by \tool.
}

\subsection{RQ4: Discovering Unreported Multiple SPs}
\label{sec:rq4}
This RQ evaluates whether \tool can discover previously unknown SPs for CVEs currently recorded with only a single patch.
We select 100 recent CVEs (all from 2026) recorded with only a single patch across all public vulnerability databases, run \tool on each, and manually verify every discovered candidate.

\noindent\textbf{Results:}
\tool identifies 48 patch candidates across 25 CVEs as potential security patches.
Manual review confirms \textbf{28 of these 48 candidates (58.3\%) as genuine security patches}, spanning \textbf{20 CVEs} (the remaining 5 CVEs yielded only false positives).
This means that 20\% of sampled CVEs have unreported additional patches, which likely underestimates the true prevalence, as follow-up patches for CVEs disclosed in early 2026 may not yet have been released.

As shown in \Cref{tab:rq4_unreported}, the vast majority (89.28\%) are fixes within the same branch, i.e., the commits that harden the vulnerability fix but are not tracked by any database.
This implies that applying only the database-recorded patch often leaves residual attack surface unaddressed, with practical consequences for organizations relying on NVD or GHSA for patch management.

\mybox{
\textbf{Finding (RQ4):} Among 100 recent CVEs, \tool discovers \textbf{28 previously unreported security patches} across 20 CVEs (20\%), demonstrating that vulnerability databases systematically under-report multi-SP phenomenon.
}

\begin{table}[t]
\centering
\caption{Unreported SP discoveries on 100 CVEs from 2026. \tool finds 28 previously unknown patches across 20 CVEs.}
\label{tab:rq4_unreported}
\begin{tabular}{lrr}
\toprule
\textbf{Multi-SP Scenario} & \textbf{Patches} & \textbf{CVEs} \\
\midrule
Same-branch fixes & 25 & 17 \\
Cross-branch fixes            &  1 &  1 \\
Cross-repository fixes            &  2 &  2 \\
\midrule
\textbf{Total}                   & \textbf{28} & \textbf{20} \\
\bottomrule
\end{tabular}
\end{table}

\section{Discussion}

\subsection{Threats to Validity}
\noindent $\bullet$ \textbf{Dataset Construction:}
\spdb is merged from four databases, so it may include incorrect SPs and omit others. Since no public dataset provides complete and accurate SPs, we mitigate this by drawing only on widely used, peer-reviewed sources.

\noindent $\bullet$ \textbf{Reliance on commits only:}
Since SPs usually appear as commits in OSS repositories~\cite{tan2021locating}, we only focused on the commit links in the references from vulnerability databases. We did not rely on the \texttt{Patch} label offered by some databases, as it is neither always available.

\noindent $\bullet$ \textbf{Reliance on GitHub only:} Given that extracting commit information from multiple platforms (e.g., GitHub, and Bitbucket) requires maintaining platform-specific API interfaces, we focus on SPs hosted on GitHub.
Based on all multi-SP patch links we collected from 2005--2026, only 3.3\% of CVEs have patches hosted exclusively outside GitHub, spanning 93 repositories on platforms such as git.kernel.org, gitlab.com, and bitbucket.org. These cases still cover the main taxonomy categories, so the exclusion is unlikely to remove an entire category of multi-SP behavior. Because linking practices may differ across platforms, however, we cannot rule out platform-specific distribution shifts.

\noindent $\bullet$ \textbf{LLM-based Labeling:} Bias of llm-based labeling for analyzing reasons in~\Cref{sec:reasons} could be inevitable. To mitigate this threat, we adopted a cross-review process and further validated a stratified random sample to confirm the reliability of the labeling, as described in~\Cref{sec:reliability_validation}. These measures improve the accuracy of commit annotation and strengthen the reliability of our conclusions.

\noindent $\bullet$ \textbf{Limitations of discovering SPs with weak observable patterns:} \tool's candidate-discovery stage requires observable evidence to bring a commit into the candidate pool before LLM determination, such as shared issue/PR/CVE references. Therefore, SPs that lack such observable candidate-discovery patterns can be missed. However, we found that 3.0\% of patch pairs in \spdb lack strong observable patterns, suggesting that these cases exist but are relatively uncommon.

\noindent $\bullet$ \textbf{Reliance on observed repository relations:} \tool's cross-repository phase uses a relation table built from \sptax{} to efficiently handle recurring repository relations, but may miss repositories with no previously observed relation. To mitigate this, \tool also falls back to fork-related repositories, and we found that disabling the table affected only 7/61 CVEs and reduced recall and precision by 5.5 and 1.2 pp, respectively, suggesting that the cross-repository results are not primarily driven by memorized repository pairs.

\subsection{Lessons Learned}

Our empirical analysis of 6,053 multi-SP CVEs across major vulnerability databases reveals several key takeaways that advance the understanding and management of SPs.

\subsubsection{Lesson 1: Missing Recorded Patches for Multi-SP CVEs Are Dangerous}
For \textit{Incomplete Fix} and \textit{Incorrect Fix} as shown in Figure~\ref{fig:sankey_taxonomy}, an unlinked corrective patch leaves downstream maintainers and tools assessing, synchronizing, or porting a fix that is already known to be inadequate~\cite{zhao2023software,zhang2024does}.
In \textit{Cross-Branch Porting}, and cross-repository categories such as \textit{Dependency Fix}, an unlinked patch hides the branch- or repository-specific fix needed in another maintenance context.
Partial patch records therefore obscure the full fixing scope, delay correct patch deployment, and can lead downstream users or tools to miss the patch that is actually applicable to their software.

\subsubsection{Lesson 2: Effective Multi-SP Tracing Needs Relation-Aware and Cross-Source Analysis}
Our tool evaluation (Section~\ref{sec:rq3_eval}) shows that the relevance between CVE descriptions and commits is not sufficient on its own but accounting for the relations among SPs across time, branches, and repositories can improve tracing effectiveness.
This suggests that future patch localization tools should leverage cross-source evidence and relation-aware signals such as temporal proximity, commit message similarity, and overlapping code changes to discover more comprehensive SPs.

\subsubsection{Lesson 3: Directions for Community and Ecosystem Enhancement}
Section~\ref{sec:db_study} shows that no single vulnerability database provides comprehensive coverage of multi-SP CVEs.
Thus, the community needs stronger coordination and standardized metadata for patch relationships.
Databases should label incomplete, incorrect, follow-up, branch-specific, and repository-specific patches to improve traceability.
Security practitioners and database maintainers can improve transparency and patch coverage by explicitly linking related commits and documenting how a vulnerability is fixed across branches or repositories.
Ranking-based tools should also balance the recall gains of larger top-$K$ values against their added manual validation cost.

\section{Related Work}

\subsection{Empirical Study of Patch Management}
Previous research has examined diverse aspects of security patching in open-source software ecosystems.
Li et al.~\cite{li2017large} analyzed patch size, latency, and regression risks at scale.
Tan et al.~\cite{tan2022understanding} studied security-patch propagation across software branches.
Ramkisoen et al.~\cite{ramkisoen2022pareco} investigated missed and duplicated patches in forked projects. Dissanayake et al.~\cite{dissanayake2022empirical} examined automation in patch management through practitioner interviews. Xie et al.~\cite{xie2024unveiling} studied post-deployment patch evolution and its impact on vulnerability-analysis tools. Woo et al.~\cite{woo2025large} assessed NVD patch effectiveness and found many incomplete or unreliable patches.
Park et al.~\cite{park2012empirical,park2017empirical} studied multi-fix bugs in general defect repair, finding that about one-quarter of bug reports require supplementary fixes, consistent with our observations on security patches.

These studies cover patch latency, propagation, effectiveness, automation, and fork coordination, but they do not systematically study the multi-SP phenomenon itself, including its prevalence and underlying causes. Our work addresses this gap and uses the resulting taxonomy to guide \tool in discovering comprehensive SP sets for a given CVE.

\subsection{Tracing Security Patches for Disclosed Vulnerabilities}
\label{sec:related_localization}
A large body of prior work studies how to identify SPs for a disclosed vulnerability, typically formulating the task as ranking candidate commits by their relevance to a CVE description.
Early methods rely on handcrafted textual, structural, or link-based signals, including PatchScout~\cite{tan2021locating}, VCMatch~\cite{wang2022vcmatch}, VFCFinder~\cite{dunlap2024vfcfinder}, Prospector~\cite{sabetta2024known}, and Tracer~\cite{xu2022tracking}.

More recent studies strengthen ranking with pretrained models, LLMs, and contextual constraints. PromVPat~\cite{zhang2024dual}, PatchFinder~\cite{li2024patchfinder}, PatchSeeker~\cite{nguyen2025patchseeker}, Taper~\cite{ran2025efficient}, and Xu et al.~\cite{xu2025revisiting} improve CVE--commit semantic matching with PLMs/LLMs, while other hybrid frameworks incorporate vulnerability metadata, repository context, and temporal cues~\cite{hommersom2024automated,nguyen2025mapping}.

Several studies go beyond identifying a single patch.
SPV~\cite{wang2025locating} applies rule-based analysis to locate branch-level or backported variants once a reference SP is known. SHIP~\cite{song2025not} is the closest to our research, as it explicitly considers the scenarios in which one CVE can correspond to multiple SPs.
However, SHIP does not explicitly investigate the multi-SP phenomenon itself, nor does it distinguish the different scenarios under which multiple SPs arise for the same vulnerability. Instead, it treats all related commits in a largely uniform manner. Therefore, the relevance of two commits is still modeled mainly through surface-level similarities, such as shared code entities and text similarity, and the final identification is ultimately resolved through ranking.

Overall, existing ranking-based approaches are not well-suited to recovering comprehensive SPs for a multi-SP CVE, because there is no principled top-$K$ cutoff that guarantees patch coverage. In contrast, our work constructs a two-level taxonomy of the multi-SP phenomenon and uses it to guide \tool: candidate-discovery patterns bound the search space without a top-$K$ cutoff, and LLM-Agent determination judges whether each candidate is a genuine SP for the target vulnerability.

\subsection{Identifying Silent Security Patches}
Prior work has also explored the identification of silent security patches, those fixing vulnerabilities that have not yet been publicly disclosed or linked to a CVE ID.

Early studies~\cite{zhou2021spi,zhou2021finding,wang2021patchrnn,wu2022enhancing,nguyen2022vulcurator} leveraged deep neural networks to learn from commit messages, code differences, or GitHub issues for automatic security patch detection, while Wang et al.~\cite{wang2019detecting,wang2020empirical} further characterized the prevalence and security implications of such secret security patches. Subsequent efforts advanced this line with richer code representations: GraphSPD~\cite{wang2023graphspd} and CoLeFunDa~\cite{zhou2023colefunda} modeled code changes as graphs to capture syntactic and semantic dependencies, and Wen et al.~\cite{wen2025repository} further scaled this idea to the repository level to account for cross-file dependencies. More recently, Tang et al.~\cite{tang2025just} and Yang et al.~\cite{yang2025code} incorporated LLMs into this task: the former augments patch representations with LLM-generated code change explanations and contrastive learning, whereas the latter enriches commit representations with development artifacts, historical vulnerability fixes, and code changes.

These studies address a fundamentally different research problem: they aim to classify arbitrary commits as security fixes in the absence of any CVE linkage, whereas our work investigates the multi-SP phenomenon for disclosed CVEs.

\section{Conclusion}
We conduct the first large-scale empirical study of the multi-SP phenomenon, revealing that multiple SPs for a single vulnerability are prevalent.
By constructing a two-level taxonomy (including 6 categories and 16 sub-categories) of reasons behind this phenomenon, we ground the design of \tool, a taxonomy-driven prototype that recovers additional SPs beyond existing patch localization tools and uncovers 28 previously unreported SPs in recently disclosed CVEs.
Our findings call for greater awareness of multi-SP vulnerabilities among database maintainers, tool builders, and security practitioners alike.

\begin{acks}
We thank the anonymous reviewers for their constructive and insightful comments, which substantially improved this paper.
This work was supported by the \grantsponsor{GS1}{National Key R\&D Program of China}{https://www.most.gov.cn/} (Grant No.~\grantnum{GS1}{2024YFE0203800}), the \grantsponsor{GS2}{Beijing-Tianjin-Hebei Natural Science Foundation Cooperation Project}{http://www.bjnsf.org/} (Grant No.~\grantnum{GS2}{25JJJJC0003}), the \grantsponsor{GS3}{Tianjin Major Science and Technology Special Project}{http://kxjs.tj.gov.cn/} (Grant No.~\grantnum{GS3}{25ZXSFSN00140}), and the \grantsponsor{GS4}{China Scholarship Council}{https://www.csc.edu.cn/} (Grant No.~\grantnum{GS4}{202506200059}).
\end{acks}

\clearpage

\section*{Data Availability Statement}
The dataset and scripts of this study are publicly available at~\cite{dataset}.

\bibliographystyle{ACM-Reference-Format}
\bibliography{myref}

\end{document}

%% file: imgs/incomplete_fix_example.tex
\begin{figure}[t]
\centering
\small
\begin{tcolorbox}[
  colback=white, colframe=black!90, boxrule=0.5pt,
  fonttitle=\small\bfseries,
  left=4pt, right=4pt, top=2pt, bottom=2pt
]

\textbf{Vulnerable Code} (\texttt{class.FroxlorInstall.php}):
\begin{lstlisting}[language=PHP, basicstyle=\ttfamily\scriptsize, showstringspaces=false, escapeinside={(*@}{@*)}]
$cmd = $mysqldump . " " . $db  (*@\hfill \textcolor{red}{\ding{55}\ unsanitized}@*)
    . " -u " . $user    (*@\hfill \textcolor{red}{\ding{55}\ unsanitized}@*)
    . " --password='" . $pass           (*@\hfill \textcolor{red}{\ding{55}\ unsanitized}@*)
    . "' --result-file=" . $file;
exec($cmd);
\end{lstlisting}

\vspace{2pt}
\tikz\draw[dashed, gray] (0,0) -- (\linewidth-8pt,0);
\vspace{2pt}

\textbf{Patch~A} (commit \texttt{Froxlor/Froxlor@62ce21c}, incomplete fix):
\begin{lstlisting}[language=PHP, basicstyle=\ttfamily\scriptsize, showstringspaces=false, escapeinside={(*@}{@*)}]
$cmd = $mysqldump . " " . escapeshellarg($db)  (*@\hfill \textcolor{green!50!black}{\ding{51}\ fixed}@*)
    . " -u " . escapeshellarg($user)    (*@\hfill \textcolor{green!50!black}{\ding{51}\ fixed}@*)
    . " --password='" . $pass           (*@\hfill \textcolor{red}{\ding{55}\ unsanitized}@*)
    . "' --result-file=" . $file;
\end{lstlisting}

\vspace{2pt}
\tikz\draw[dashed, gray] (0,0) -- (\linewidth-8pt,0);
\vspace{2pt}

\textbf{Patch~B} (commit \texttt{Froxlor/Froxlor@7e36127}, follow-up):
\begin{lstlisting}[language=PHP, basicstyle=\ttfamily\scriptsize, showstringspaces=false, escapeinside={(*@}{@*)}]
$cmd = $mysqldump . " " . escapeshellarg($db)  (*@\hfill \textcolor{green!50!black}{\ding{51}\ fixed}@*)
    . " -u " . escapeshellarg($user)    (*@\hfill \textcolor{green!50!black}{\ding{51}\ fixed}@*)
    . " --password='" . escapeshellarg($pass)  (*@\hfill \textcolor{green!50!black}{\ding{51}\ fixed}@*)
    . "' --result-file=" . $file;
\end{lstlisting}

\end{tcolorbox}

\caption{An example of an incomplete fix (CVE-2020-10235).}
\label{fig:incomplete-fix-example}
\end{figure}

%% file: imgs/cross_branch_example.tex
\definecolor{diffgreen}{RGB}{230,255,230}
\definecolor{diffred}{RGB}{255,230,230}
\definecolor{commentgray}{RGB}{120,120,120}

\newcommand{\codeadd}[1]{\colorbox{diffgreen}{\makebox[\dimexpr\linewidth-8pt\relax][l]{\ttfamily\scriptsize #1}}}
\newcommand{\codedel}[1]{\colorbox{diffred}{\makebox[\dimexpr\linewidth-8pt\relax][l]{\ttfamily\scriptsize #1}}}
\newcommand{\codecmt}[1]{{\ttfamily\scriptsize\textcolor{commentgray}{#1}}}
\newcommand{\codeomit}[1]{{\ttfamily\scriptsize\textcolor{commentgray}{\quad\ldots\quad// #1}}}

\begin{figure}[t]
\centering
\small
\begin{tcolorbox}[
    colback=white, colframe=black!90, boxrule=0.5pt,
  fonttitle=\small\bfseries,
  left=4pt, right=4pt, top=2pt, bottom=2pt
]

\textbf{Patch~A} - branch \texttt{4.x} (commit \texttt{d541378}, minimal fix):\par\vspace{2pt}
\codecmt{// lib/handlebars/helpers/lookup.js}\par\nobreak
\codedel{- if (field === 'constructor' \&\& ...) \{}\par\nobreak
\codeadd{+ if (String(field) === 'constructor' \&\& ...) \{}

\vspace{4pt}
\tikz\draw[dashed, gray] (0,0) -- (\linewidth-8pt,0);
\vspace{4pt}

\textbf{Patch~B} - branch \texttt{3.x} (commit \texttt{156061e}):\par\vspace{2pt}
\codeomit{utils.js: define dangerousPropertyRegex blocklist}\par\vspace{3pt}
\codecmt{// base.js - broader guard replaces constructor-only check}\par\nobreak
\codedel{- if (field === 'constructor'}\par\nobreak
\codedel{-\qquad \&\& !obj.propertyIsEnumerable(field)) \{}\par\nobreak
\codeadd{+ if (dangerousPropertyRegex.test(String(field))}\par\nobreak
\codeadd{+\qquad \&\& !Object.prototype.hasOwnProperty.call(obj, field)) \{}\par\vspace{3pt}
\codeomit{javascript-compiler.js: add dangerousPropertyRegex guard}\par

\end{tcolorbox}

\caption{Cross-branch porting: CVE-2019-20920 (handlebars.js).}
\label{fig:cross-branch-example}
\end{figure}